\documentstyle[epsf]{l-aa}
\begin{document}

\thesaurus{03(11.04.1;   
              11.05.1;   
              11.06.2;   
              11.16.1)}  

\title{Is the shape of the luminosity profile of dwarf elliptical
galaxies a useful distance indicator?}

\author{Bruno Binggeli\,\inst{1} 
\and Helmut Jerjen\,\inst{2}} 

\offprints{B.~Binggeli, e-mail : binggeli@astro.unibas.ch}   

\institute{Astronomisches Institut, Universit\"at Basel, Venusstrasse 7,
CH-4102 Binningen, Switzerland
\and
Mount Stromlo and Siding Spring Observatories, Private Bag, Weston Creek
PO,
ACT\,2611, Canberra, Australia} 

\date{Received  ; accepted }
\maketitle
\markboth{Binggeli and Jerjen: dEs as distance indicators}
{Binggeli and Jerjen: dEs as distance indicators}  

\begin{abstract}
The shape of the surface brightness profile of dE galaxies,
quantified by parameter $n$ of S\'ersic's generalized profile law,
has recently been put forward as new extragalactic distance indicator
(Young \& Currie 1994). Its  
application to the Virgo cluster has subsequently led to the claim that the
Virgo dEs are not lying in the cluster core but are
distributed in a prolate structure stretching from 8 to 20
Mpc distance (Young \& Currie 1995).

This claim is refuted here.
We have fitted a S\'ersic law to 
the surface brightness profiles of 128 Virgo cluster dEs and
dS0s from the photometry of Binggeli \& Cameron 
(1991). The dispersion of the $n - M$ relation is indeed large
($\sigma_{\rm rms} \approx$ 0.9 mag). 
However, we argue that this scatter is not due to the depth of the Virgo 
cluster, but is essentially 
intrinsic. Contrary to what one would expect from
the cluster depth hypothesis,
there is no clear velocity-``distance'' relation for a
sample of 43 Virgo dEs and dS0s with known redshifts.
The analysis of Young \& Currie (1995) is hampered by the
use of low-resolution photometry and flawed by the assumption that
the $n - M$ and $n - R$ relations can be used {\em independently}.

By combining different S\'ersic law parameters, the scatter of the
scaling relations can be 
reduced somewhat, but never below $\sigma_{\rm rms} 
\approx$ 0.7 mag, at least for the Virgo cluster. For the purpose of
distance measurements, this falls
short of the well-established Tully-Fisher and $D_{\rm n} - \sigma$
methods, and it is comparable to what one can get already from the 
$\langle \mu \rangle_{\rm eff} - M$ relation for dEs, which does
not require any profile modelling.
 
\end{abstract}

\section{Introduction}
The surface brightness profile of dwarf elliptical (dE) galaxies
follows a 3-fold trend: with increasing galaxy luminosity (1) the 
mean surface brightness increases, (2) the profile becomes flatter in the
outer region, while (3) the profile becomes more cuspy, i.e.~more E-like in 
the core region (Binggeli \& Cameron 1991). The two latter 
characteristics, combined in the overall {\it shape}, or {\it curvature}\/ 
of the dE profile, can be quantified and 
parametrized by means of S\'ersic's (1968) $r^{n}$ law for the radial surface 
brightness profile of galaxies, which is a 
simple generalization of de\,Vaucouleurs' 
$r^{1/4}$ and exponential laws (Davies et al.~1988, 
Young \& Currie 1994, Jerjen 1995). 
The profile curvature of dEs is described by S\'ersic's exponent $n$. 

Working with a small sample of Fornax cluster dEs, Young \& Currie 
(1994, hereafter 
YC94) found 
that the correlation between $n$ and total magnitude of the galaxy
was so tight, with 
an rms scatter of only $\approx$ 0.47 mag, that it could be used 
to derive a distance of the Fornax cluster, based on a handful of local, 
calibrating
dEs. Hence the claim of a ``new extragalactic distance indicator''.

In a second paper, Young \& Currie (1995, hereafter YC95) applied a 
variant of this 
new distance indicator, viz.~the relation between $n$ and the 
logarithm of S\'ersic's 
scale length $r_0$ (see below), to a sample of 64 Virgo cluster dEs, 
for which they
had derived profiles from low-resolution Schmidt plates. 
The scatter of the $n - \log r_0$ 
relation turned out to be much larger for these Virgo cluster dEs 
than for an external 
sample of Local Group and Fornax cluster dwarfs. 
Nevertheless, YC95 argued, by way of 
comparing the results from the apparently independent 
$n - M$ and $n - \log r_0$ relations, 
that the {\em intrinsic}\/ scatter for Virgo 
cluster dEs would be equally small. The 
inevitable conclusion on this assumption was that the large 
observed scatter for Virgo cluster dEs must be attributed to the {\em depth}\/
of the Virgo cluster. Moreover, the resulting filamentary cloud of dEs, 
stretching from ca.~8 to 20 Mpc in Young and Currie's distance scale, and
apparently by chance (?) being aligned with our line of sight, is in perfect
accord with the filament of spiral galaxies advocated by Fukugita 
et al.~(1993) and Yasuda et al.~(1997) based on Tully-Fisher distances.

This we found alarming. Spiral galaxies are supposed to avoid the core of
the Virgo cluster (Binggeli et al.~1987). However, from all we know
(cf.~Ferguson \& Binggeli 1994 for a recent review, Stein et al.~1997) 
-- dwarf ellipticals, like ellipticals in general, reside only in dense galaxy 
environments, which would exclude loosely bound 
clouds or filaments of galaxies.

Motivated by this contradiction, we went back to the Virgo cluster photometry
of Binggeli \& Cameron (1991, 1993, hereafter BC91 and BC93) and selected 128 
highly resolved dE and dS0 profiles that are 
well explained by S\'ersic's 
generalized law. This sample, as well as the procedure and the results of the 
fitting are presented in Sect.~2. 
Our analysis of the data, in Sect.~3, confirms that
the scatter of either the $n - M$ or the $n - \log r_0$ relation is very large.
It cannot be reduced to below ca.~0.7 mag, which is comparable to the scatter
of the surface brightness - luminosity ($\langle \mu \rangle_{\rm eff} - M$) 
relation (BC91). This is 
simply too large for a distance indicator 
to be useful -- {\em if}\/ the scatter is intrinsic.

As we argue in Sect.~4, there is indeed some evidence that this large scatter
for Virgo {\em is}\/ intrinsic, i.e.~cosmic. The residuals from the
$n - M$ and $n - \log r_0$ relations do not correlate with the velocities
of the dwarfs -- which they should if these galaxies were distributed in a 
filament outside the Virgo cluster core. We also discuss why we think
Young \& Currie's analysis is flawed. 
Our sober conclusions are given in Sect.~5.

\section{Generalized profile parameters of Virgo cluster dwarfs}
Following the notation of YC94, S\'ersic's (1968) generalized profile law can 
be written as
\begin{equation}
\sigma(r) = \sigma_0\:e^{-{(r/r_0)}^{n}}\,\,,
\end{equation}
where $\sigma$ is the surface brightness (light intensity per area) at the
mean galactocentric radius $r$. There are three parameters: (1) the central
surface brightness $\sigma_0$, (2) the characteristic radius, or scale length
$r_0$ at which $\sigma = \sigma_0/e$, and (3) 
the shape parameter $n$. The corresponding 
law in the magnitude (logarithmic) representation is
\begin{eqnarray}
\mu(r) & = & -2.5 \log \sigma(r) + const \nonumber\\
       & = & \mu_0 + 1.086\,{(r/r_0)}^{n}\,\,,
\end{eqnarray}
with $\mu_0 = -2.5 \log \sigma_0 + const$. The quantity $\mu$ is again called
``surface brightness'' but has now the conventional unit of 
(mag arcsec$^{-2}$). For $n = 1$ S\'ersic's law is identical to the 
exponential; for $n = 1/4$ it reduces to de\,Vaucouleurs' (1948) $r^{1/4}$
law. In fact, S\'ersic (1968) expressed his law in terms of de\,Vaucouleurs'
parameters ($\mu_{\rm eff}, r_{\rm eff}$), which can easily be transformed
to ($\mu_0, r_0$) by generalizing de\,Vaucouleurs' exponent 1/4 to 1/$n$;
hence $n_{\rm S\acute{e}rsic} = 1/n_{\rm here}$.

It has become clear in recent years that S\'ersic's (1968) generalization
of de\,Vaucouleurs' law is fully explaining the variety of luminosity
profiles of normal elliptical galaxies (Caon et al.~1993, D'Onofrio et 
al.~1994). As we now realize, it also provides an excellent description
of the profiles of {\it dwarf}\/ ellipticals. Unfortunately, BC91
were not aware of this law (in spite of Davis et al.~1988)
but fitted exponentials and King (1966) models to their large ({\small $N$}
$\approx$ 200) 
and homogeneous set of well-resolved light profiles of early-type 
Virgo dwarfs. 
Indeed, the exponential and King model fits were not satisfactory 
for bright dwarfs. 
There was always what was called an ``extended, central luminosity 
excess''. S\'ersic's 
law nicely takes care of these excesses, i.e.~with a S\'ersic law 
fit there is no excess 
left over -- except that caused by an unresolved, quasi-stellar 
central nucleus, 
if present. Furthermore, the central light deficiency observed 
for dwarfs fainter than $M_{B_T}\la-16$ 
is best approximated with S\'ersic's parameter 
$n$ being $>1$.

For the present investigation we have fitted S\'ersic laws to the (unpublished)
dwarf profiles of BC91/93. In contrast to YC94, who modelled their 
{\em differential}\/ profiles with Eq.(2) by linear regression, we fitted our
{\em growth curves}\/ 
(cumulative intensity profiles) by a $\chi_{\rm min}^{2}$ 
method with the corresponding cumulative S\'ersic law, which is (Jerjen 1995):
\begin{eqnarray}
I(r) & = & \int^r_0\!\sigma(R)\,2\pi R\,dR \nonumber\\
     & = & \frac{2 \pi \sigma_0 r_0^2}{n} \cdot \gamma [2/n, {(r/r_0)}^n]\,\,,
\end{eqnarray}
where $\gamma [a, x] = \int^x_0 exp(-t)\,t^{a-1} dt$ is the Incomplete Gamma
function. The {\em total}\/ model intensity is then:
\begin{equation}
I_{\rm T} = I(\infty) = \frac{2 \pi \sigma_0 r_0^2}{n} \cdot \Gamma [2/n]\,\,,
\end{equation}
with $\Gamma$ as the well-known Gamma function, or -- in terms of
magnitudes --
\begin{equation}
m_{\rm T} = \mu_0 - 5 \log r_0 + 2.5 \log (n/\Gamma[2/n]) - 2.00\,\,,
\end{equation}
which can be compared with the corresponding {\em measured}\/ $B_{\rm T}$
value to test the goodness of a fit.

The innermost 3$\arcsec$ of 
the galaxy profiles were excluded from the fitting in 
order to avoid the central, semi-stellar 
nuclei of the dwarfs classified dE,N (or dS0,N). Likewise, 
an outer limiting radius for the fitting was set at a surface
brightness level of 27.0 B\,arcsec$^{-2}$. 

Our sample drawn form BC93 was restricted to dwarfs brighter than
$B_{\rm T}^{\rm lim}$ = 18, which
had also been the limit of completeness for the
surveyed cluster region (cf.~BC91). This left 158 objects. Ten of these
were excluded because their modelled total magnitude, calculated with Eq.(5),
differed from the measured $B_{\rm T}$ by more than 1 mag, the reason 
for the discrepancy being due to the presence of a very strong central
nucleus or some irregularity in the outer profile. Aside from these cases,
the mean and standard deviation of the difference between model and observed
total magnitude is (model {\em minus} observed) 0.05 $\pm$ 0.26, which clearly
shows the goodness and appropriateness of the S\'ersic law.

Although we think that the growth curve fitting is more reliable than
the differential profile fitting, we have performed the latter as well,
as a test for consistency. Mean and standard deviation of the difference
in the S\'ersic parameters, determined in both ways (growth curve {\em minus}
differential), amount to $\Delta n = - 0.003 \pm 0.16$, 
$\Delta (\log r_0) = - 0.04 \pm 0.40$ ($r$ in arcsec), and $\Delta \mu_0 = 
- 0.09 \pm 0.59$ B arcsec$^{-2}$.
Clearly, $n$ is the most stable parameter of the S\'ersic law. The scatter
in $\Delta (\log r_0)$, however, is surprisingly large, the implications of 
which will be discussed further below.

For a further purification of our sample -- while keeping it 
representative and fairly complete -- , we have excluded 20 dwarfs with
$|\Delta n|$ (growth curve $-$ differential) $>$ 0.5, leaving us with
a final sample of 128 objects. The best-fitting (growth curve) S\'ersic law 
parameters ($\mu_0$, $\log r_0$, $n$) of these 128 dwarfs are 
listed along with 
Virgo Cluster Catalog number (VCC, Binggeli et al.~1985), 
morphological type, 
total $B$\,magnitude, and heliocentric 
velocity (if available) in Table 1. Fig.~1 
shows a few representative 
dwarf profiles and the corresponding S\'ersic law fits. 
 
%
\begin{table*}
\caption[]{S\'ersic profile parameters of 128 Virgo cluster dEs and dS0s}
\footnotesize
\begin{tabular}{rlccrcr|rlccrcr}
\hline
\vspace{-4mm}\\
VCC & Type & $B_{\rm T}$ & $n$ & $\log r_0$ & $\mu_0$ & 
$V_{\rm hel}$ & VCC & Type & $B_{\rm T}$ & n & $\log r_0$ & $\mu_0$
& $V_{\rm hel}$ \\
& & & & {\scriptsize (\,\arcsec\,)} & {\scriptsize (B/$\sq\arcsec$)} 
& {\scriptsize (km/s)} 
& & & & & {\scriptsize (\,\arcsec\,)} & {\scriptsize (B/$\sq\arcsec$)} 
& {\scriptsize (km/s)}\\[-0.5mm] 
\hline
\vspace{-4mm}\\
109 & dE,N & 16.06 & 0.88 & 0.75 & 22.06 & & 1212 & dE,N & 16.94 & 1.03 &
1.27 & 25.07 & \\[-0.5mm]
168 & dE & 17.10 & 0.81 & 0.63 & 22.57 & 682 & 1213 & dE,N & 16.42 & 0.77 & 
0.74 & 22.69 & \\[-0.5mm]
170 & dS0 & 14.56 & 0.47 & 0.13 & 20.31 & 1493 & 1254 & dE,N & 15.51 & 0.84 &
0.77 & 21.77 & 1350\\[-0.5mm]
235 & dE,N & 16.87 & 1.54 & 1.16 & 24.16 & & 1261 & dE,N & 13.56 & 0.55 &
0.38 & 19.59 & 1850\\[-0.5mm]
299 & dE & 17.31 & 0.85 & 0.64 & 22.87 & & 1264 & dE,N & 17.31 & 0.98 & 0.88 &
23.62 & \\[-0.5mm]
354 & dE & 16.60 & 1.23 & 1.06 & 23.59 & & 1268 & dE,N & 17.24 & 1.82 & 1.30 &
25.30 & \\[-0.5mm]
389 & dS0,N & 14.21 & 0.35 & --0.82 & 17.81 & 1330 & 1308 & dE,N & 15.64 &
0.54 & --0.28 & 19.17 & 1721\\[-0.5mm]
444 & dE & 17.22 & 1.47 & 1.16 & 24.70 & & 1348 & dE,N & 15.87 & 0.65 & 0.64 &
21.38 & 1679\\[-0.5mm]
490 & dS0,N & 14.05 & 0.77 & 0.91 & 21.31 & 1293 & 1353 & dE,N & 16.61 & 0.80
& 0.42 & 21.27 & \\[-0.5mm]
494 & dE & 16.64 & 0.98 & 0.82 & 23.20 & & 1389 & dE,N & 15.91 & 0.54 & 0.02 &
20.18 & \\[-0.5mm]
510 & dE,N & 15.13 & 0.60 & 0.51 & 21.27 & & 1392 & dS0,N & 14.86 & 0.81 & 0.93
& 22.09 & \\[-0,5mm]
543 & dE & 14.39 & 0.37 & --0.85 & 18.11 & 861 & 1396 & dE,N & 17.21 & 0.80 &
0.89 & 23.95 & \\[-0.5mm]
554 & dE,N & 17.11 & 1.00 & 1.00 & 23.95 & & 1399 & dE,N & 16.49 & 0.40
& --0.68 &
19.90 & \\[-0.5mm]
594 & dE & 17.13 & 1.50 & 1.05 & 24.15 & & 1407 & dE,N & 15.49 & 0.46 & 
--0.28 & 19.29 & 941 \\[-0.5mm]
608 & dE,N & 14.94 & 0.63 & 0.47 & 20.75 & 1803 & 1417 & dE & 15.76 & 0.58 &
0.18 & 21.06 & \\[-0.5mm]
684 & dE,N & 16.04 & 0.70 & 0.48 & 21.41 & & 1420 & dE,N & 16.41 & 0.50 & 
--0.28 & 19.96 & 1022 \\[-0.5mm]
753 & dE,N & 16.37 & 0.98 & 1.01 & 23.38 & & 1432 & dE & 17.10 & 1.44 & 0.94
& 23.46 & \\[-0.5mm]
761 & dE & 17.26 & 0.67 & 0.66 & 23.53 & & 1444 & dE,N & 16.05 & 0.54 & --0.05
& 20.50 & \\[-0.5mm]
765 & dE,N & 16.49 & 0.80 & 0.35 & 20.80 & & 1446 & dE,N & 16.00 & 0.77 &
0.69 & 22.02 & \\[-0.5mm]
769 & dE & 17.24 & 0.93 & 0.59 & 22.43 & & 1451 & dE,N & 16.47 & 0.54 & --0.16
& 20.37 & \\[-0.5mm]
779 & dE,N & 17.67 & 1.24 & 1.01 & 24.42 & & 1491 & dE,N & 15.24 & 0.52 & 
--0.06 & 19.23 & 1903\\[-0.5mm]
781 & dS0,N & 14.72 & 0.54 & 0.00 & 19.10 & --254 & 1496 & dE,N & 17.92 &
0.85 & 0.52 & 22.88 & \\[-0.5mm]
810 & dE,N & 16.95 & 1.02 & 0.77 & 22.73 & --340 & 1509 & dE,N & 16.42 & 0.84 
& 0.78 & 22.71 & \\[-0.5mm]
812 & dE,N & 17.03 & 0.85 & 0.73 & 22.99 & & 1523 & dE,N & 17.64 & 1.07 & 0.75 
& 23.40 & \\[-0.5mm]
815 & dE,N & 16.10 & 0.56 & 0.26 & 21.23 & --700 & 1539 & dE,N & 15.68 & 
0.89 & 0.86 & 22.50 & 1390\\[-0.5mm]
823 & dE,N & 16.06 & 0.86 & 0.68 & 21.78 & 1691 & 1553 & dE & 16.69 & 1.05 &
1.03 & 23.64 & \\[-0.5mm]
856 & dE,N & 14.25 & 0.66 & 0.57 & 20.36 & 972 & 1561 & dE,N & 15.82 & 0.97 &
1.17 & 23.78 & \\[-0.5mm]
870 & dS0,N & 15.52 & 0.42 & --0.68 & 18.77 & 1277 & 1563 & dE,N & 16.11 & 
0.87 & 0.89 & 22.95 & \\[-0.5mm]
871 & dE,N & 15.79 & 0.72 & 0.71 & 22.22 & & 1567 & dS0,N & 14.52 & 0.63 & 
0.67 & 21.24 & 1440\\[-0.5mm]
872 & dE,N & 17.00 & 0.83 & 0.63 & 22.55 & & 1577 & dE & 16.14 & 0.61 & 0.32 
& 21.29 & \\[-0.5mm]
896 & dE,N & 17.96 & 1.20 & 0.80 & 23.61 & & 1622 & dE & 17.87 & 2.47 & 1.17
& 25.06 & \\[-0.5mm]
916 & dE,N & 16.04 & 0.53 & --0.27 & 18.83 & 1349 & 1629 & dE & 17.27 & 0.67 &
0.16 & 21.86 & \\[-0.5mm]
926 & dE & 16.97 & 0.41 & --0.44 & 21.07 & & 1661 & dE,N & 15.97 & 0.80 &
0.88 & 22.75 & 1400\\[-0.5mm]
929 & dE,N & 13.82 & 0.55 & 0.30 & 19.39 & 910 & 1688 & dE & 17.59 & 1.49 &
0.83 & 23.30 & \\[-0.5mm]
931 & dE,N & 16.43 & 1.17 & 1.00 & 23.25 & & 1704 & dE & 15.79 & 0.49 & --0.12
& 20.28 & \\[-0.5mm]
933 & dE,N & 16.60 & 0.69 & 0.45 & 22.52 & & 1711 & dE,N & 16.48 & 0.85 & 0.68
& 22.27 & \\[-0.5mm]
936 & dE,N & 15.81 & 0.60 & 0.11 & 20.67 & & 1717 & dE & 16.50 & 0.91 & 0.96 & 
24.22 & \\[-0.5mm]
940 & dE,N & 14.72 & 0.80 & 0.83 & 21.45 & 1563 & 1732 & dE & 17.77 & 0.75 &
0.59 & 23.22 & \\[-0.5mm]
949 & dE,N & 15.48 & 0.59 & 0.47 & 21.49 & & 1743 & dE & 15.50 & 0.52 & --0.04
& 20.21 & 1279\\[-0.5mm]
951 & dE,N & 14.35 & 0.50 & 0.16 & 19.87 & 2066 & 1745 & dE & 17.44 & 0.60 &
0.54 & 23.24 & \\[-0.5mm]
974 & dE,N & 16.11 & 0.68 & 0.35 & 21.34 & & 1762 & dE & 16.46 & 0.80 & 0.48 &
21.47 & \\[-0.5mm]
992 & dE,N & 16.81 & 0.87 & 0.73 & 22.62 & & 1767 & dE,N & 16.45 & 0.67 & 0.53
& 22.19 & \\[-0.5mm]
1039 & dE & 17.11 & 0.83 & 0.60 & 22.60 & & 1773 & dE,N & 16.16 & 0.70 & 0.65
& 22.32 & \\[-0.5mm]
1044 & dE,N & 16.98 & 0.88 & 0.59 & 22.25 & & 1779 & dS0 & 14.83 & 0.45 &
--0.29 & 19.02 & 1226\\[-0.5mm]
1075 & dE,N & 15.08 & 0.46 & 0.00 & 20.21 & 1844 & 1796 & dE,N & 16.52 & 0.65 
& 0.46 & 22.04 & \\[-0.5mm]
1076 & dE,N & 17.36 & 1.24 & 1.04 & 24.19 & & 1812 & dE,N & 17.78 & 0.88 & 
0.61 & 22.79 & \\[-0.5mm]
1087 & dE,N & 14.31 & 0.54 & 0.30 & 19.98 & 645 & 1826 & dE,N & 15.70 & 0.68 &
0.21 & 19.97 & 2033\\[-0.5mm]
1092 & dE,N & 17.05 & 1.07 & 0.85 & 23.17 & & 1828 & dE,N & 15.33 & 0.45 & 
--0.15 & 19.95 & 1517\\[-0.5mm]
1093 & dE,N & 16.85 & 1.04 & 0.94 & 23.52 & & 1857 & dE & 15.07 & 1.01 & 1.14 
& 22.67 & 634\\[-0.5mm]
1095 & dE,N & 17.93 & 0.48 & --0.38 & 21.54 & & 1874 & dE & 17.68 & 0.77 &
0.75 & 24.06 & \\[-0.5mm]
1099 & dE,N & 17.71 & 0.86 & 0.53 & 22.69 & & 1876 & dE,N & 15.05 & 0.72 & 
0.64 & 21.09 & 45\\[-0.5mm]
1101 & dE,N & 15.78 & 0.55 & 0.32 & 21.47 & & 1886 & dE,N & 15.49 & 0.77 &
0.65 & 21.40 & 1159\\[-0.5mm]
1104 & dE,N & 15.22 & 0.52 & 0.04 & 19.83 & 1704 & 1895 & dE & 14.91 & 0.43 &
--0.39 & 18.70 & 1032\\[-0.5mm]
1115 & dE,N & 17.69 & 1.58 & 1.09 & 24.61 & & 1896 & dS0,N & 14.82 & 0.80 &
0.76 & 21.13 & 1731\\[-0.5mm]
1119 & dE,N & 17.36 & 1.01 & 0.78 & 23.25 & & 1915 & dE & 17.13 & 0.96 & 0.87
& 23.51 & \\[-0.5mm]
1120 & dE,N & 17.17 & 0.76 & 0.54 & 22.54 & & 1936 & dS0,N & 15.68 & 0.49 &
--0.34 & 19.36 & \\[-0.5mm]
1122 & dE,N & 14.60 & 0.37 & --0.80 & 17.90 & 436 & 1942 & dE,N & 16.77 & 1.09
& 0.91 & 23.15 & \\[-0.5mm]
1149 & dE & 17.44 & 1.08 & 1.23 & 25.28 & & 1949 & dS0,N & 14.38 & 0.60 & 0.55
& 20.79 & 2077\\[-0.5mm]
1167 & dE,N & 15.91 & 0.69 & 0.60 & 21.88 & & 2029 & dE & 17.79 & 0.87 & 0.51
& 22.74 & \\[-0.5mm]
1172 & dE,N & 16.23 & 0.72 & 0.38 & 21.03 & & 2042 & dE,N & 14.84 & 0.71 & 
0.86 & 21.96 & \\[-0.5mm]
1183 & dS0,N & 14.37 & 0.41 & --0.55 & 18.36 & 1387 & 2043 & dE & 17.94 & 0.84
& 0.66 & 23.58 & \\[-0.5mm]
1185 & dE,N & 15.68 & 0.66 & 0.53 & 21.56 & 500 & 2045 & dE,N & 16.33 & 0.78 
& 0.56 & 21.74 & \\[-0.5mm]
1207 & dE,N & 17.55 & 0.92 & 0.82 & 23.72 & & 2048 & dS0 & 13.81 & 0.29 & 
--1.52 & 16.59 & 1095\\[-0.5mm]
1209 & dE & 17.80 & 1.05 & 0.62 & 22.89 & & 2054 & dE & 16.68 & 0.36 & --0.80
& 20.14 & \\ 
\vspace{-4mm}\\ 
\hline\\
\end{tabular}
\normalsize
\end{table*}
%
%
%

\begin{figure}
\begin{center}
\epsfxsize 85mm
\epsffile{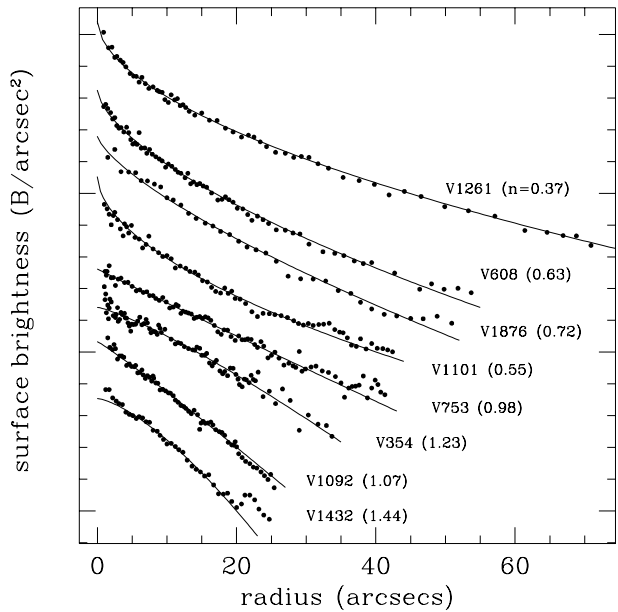}
\caption[]{Mean radial surface brightness profiles of 8 representative
Virgo cluster dEs. The profiles are ordered by total magnitude (brightest 
object on top) but are shifted by arbitrary amounts along the ordinate. The 
observed 
profiles are given by the dots. The lines are best-fitting S\'ersic 
profiles whose shape parameter $n$ is given in parentheses behind
the name of the galaxy. Note the systematic trend of the curvature $n$.}
\end{center}
\end{figure}

The accuracy of the best-fitting parameters is most reliably evaluated by a
comparison with external data. 
Unfortunately, a detailed, galaxy-to-galaxy comparison with YC95 is not useful
for this purpose. Due to the low resolution of Schmidt plates,
the radial profiles of YC95 are flattened, 
i.e.~smoothed out compared with ours.
Indeed, for the 30 galaxies in common we find $n_{\rm YC} - n_{\rm here}
\approx 1.3 \pm 0.25 (1\,\sigma)$.
Durrell (1997), whose results were published
during this 
writing, provides high-quality data for 13 Virgo dEs, seven of which
are in common with the present sample. A comparison of our S\'ersic law
parameters yields an rms (1$\sigma$) scatter, i.e.~assumed error,
for either Durrell (1997) or
the present 
analysis, of 0.10 in $n$, 0.55 B arcsec$^{-2}$ 
in $\mu_0$, and 0.32 in $\log r_0$ ($r$ in arcsec). 
Thus the agreement is excellent for the shape parameter $n$,
which indeed seems to be the most robust and reliable 
profile parameter. On the other hand, the scatter in $\log r_0$ is again
quite large. Surprisingly, it does not get smaller if we use  
differential profile fits (as did Durrell) -- but here,
part of the problem might be the presence of  
colour gradients in the dwarf profiles,
as Durrell's (1997) data are, essentially, in the R band, ours in B.

Fig.~2 shows the $n - B_{\rm T}$ relation for our dwarf sample, complemented
by Caon et al.'s (1993) data for normal (``giant'') E and S0 galaxies, which 
we have restricted to their ``good'' and ``fair'' quality fits, and by a few
compact, M32-type ellipticals from BC93, for which we have derived S\'ersic 
law parameters in the same way as for the dwarfs. The {\em absolute}\/
magnitude scale is $M_{{\rm B}_{\rm T}} = B_{\rm T} - 31.5$, assuming 
$D_{\rm Virgo}$ = 20 Mpc. 
\begin{figure}
\begin{center}
\epsfxsize 80mm
\epsffile{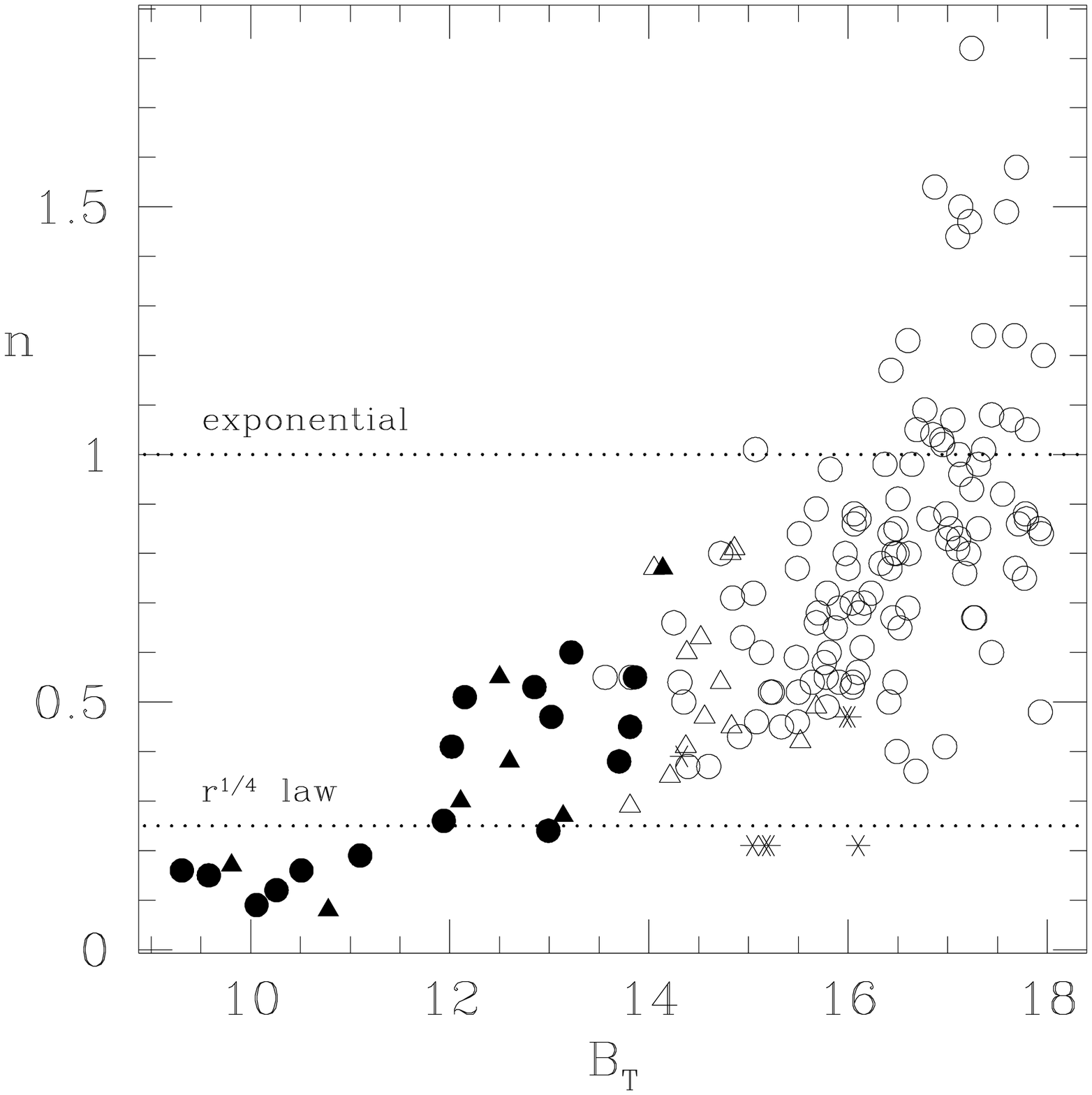}
\caption[]{S\'ersic's shape parameter $n$ versus total blue magnitude for
early-type galaxies in the Virgo cluster. Dwarf galaxies (open circles: dE,
open triangles: dS0) are from the present work. Data for Es (filled circles)
and S0s (filled triangles) are taken from Caon et al.~(1993). A few compact,
M32-type Es from BC93 are added as asterisks. Note the continuity of $n$ as 
opposed to the traditional 
bimodality (exponential versus $r^{1/4}$ law), which 
is indicated by the dotted lines. }
\end{center}
\end{figure}

Two important points are evident from Fig.~2. (1) There is indeed a clear
$n - M$ relation for dwarf ellipticals which {\em might}\/ 
be used as distance 
indicator. This is the topic of the present paper. (2) There is a 
{\em continuity}
in the profile shape between normal and dwarf ellipticals (and S0s),
with the exception of the compact, M32-type Es. In fact, this continuity 
holds for all three S\'ersic profile parameters, not just
for $n$ (Jerjen \& Binggeli 1997). This is quite surprising, because until 
recently the emphasis was rather on the {\em dis}\/continuity between Es 
and dEs in the core parameters (e.g., BC91). As it appears now, that 
discontinuity 
is restricted to the very central part of the galaxies (with 
galactocentric radius $r < 3\arcsec$, i.e.~$r <$ 
300 pc if $D_{\rm Virgo}$ = 20 Mpc). 
The overall similarity of their light distribution strongly suggests 
a common formation mechanism for normal and dwarf ellipticals (but 
probably excluding dwarf spheroidals). For a further discussion of this 
aspect, the reader is referred to Jerjen \& Binggeli (1997).   

\section{Correlation analysis}
In the following we explore the correlations of S\'ersic's profile parameters 
with total magnitude for our sample of 128 Virgo dEs and dS0s, focussing on 
possible applications to distance measurements. As seen in Fig.~2, the $n - M$ 
relation is not linear at faint magnitudes. 
We therefore propose to use $\log n$ 
instead of $n$. 
Fig.~3 shows that the $\log n - M$ relation indeed appears to be 
linear, i.e.~is compatible with the assumption of linearity. In this and 
the following figures 
we distinguish between different dwarf subtypes by plotting 
dE, dE,N, dS0, and dS0,N systems with 
different symbols. However, there is no systematic
difference with respect to type, other than the well-known tendency of 
nucleated dwarfs to be brighter than the non-nucleated ones.

The $\log n - M$, as the $n -M$ relation for Virgo dwarfs shows a large scatter
(Fig.~3), in accord with YC95. 
But in contrast to these authors we will attribute 
only a small part of this large scatter to the depth of the Virgo cluster. A 
linear regression for log $n$ versus $B_{\rm T}$ gives 
\begin{equation}
B_{\rm T} = 4.610 \log n + 16.844\,\,,
\end{equation}
which is shown as line in Fig.~3. The fitting was restricted to 
objects with $\log n \le$ 0. This should approximately 
account for the magnitude cut-off
at $B_{\rm T}$ = 18, which otherwise would artificially reduce the scatter.
Note that $B_{\rm T}$ is chosen as the dependent variable because we are
interested in $\log n$ as distance indicator. The rms (1$\sigma$) scatter
of $B_{\rm T}$ around this line amounts to a large $\sigma_{\rm obs}$ = 
0.92 mag.

Had we used differential profile fits instead of growth curve fits, the
$B_{\rm T} - \log n$ relation would only slightly differ from Eq.(6),
with a marginally larger $\sigma_{\rm obs}$ = 0.98, in accord with the
small $\sigma (\Delta n)$ quoted in Sect.~2.

If we allow for a photometric error in $B_{\rm T}$ of $\sigma_{\rm phot}
\simeq$ 0.2 mag and an error of 0.08 in $\log n$ that propagates via Eq.(5) to
$\sigma_{\rm profile}$ = 0.33 mag , both reckoned from a comparison with 
Durrell (1997), as well as a {\em conventional}\/ Virgo cluster depth 
in terms of magnitudes of $\sigma_{\rm depth} \simeq$ 0.2 mag (cf.~Binggeli et 
al.~1987), we arrive at a still very large {\em intrinsic} scatter of
\begin{equation}
\sigma_{\rm intr} = \sqrt[]{\sigma^2_{\rm obs} - \sigma^2_{\rm phot} 
  - \sigma^2_{\rm profile} - \sigma^2_{\rm depth}} = 0.81\,\,{\rm mag}\,\,.
\end{equation}
Clearly, this relation is of no use for distance measurements, even if we
could control, or significantly reduce the errors.

\begin{figure}
\begin{center}
\epsfxsize 75mm
\epsffile{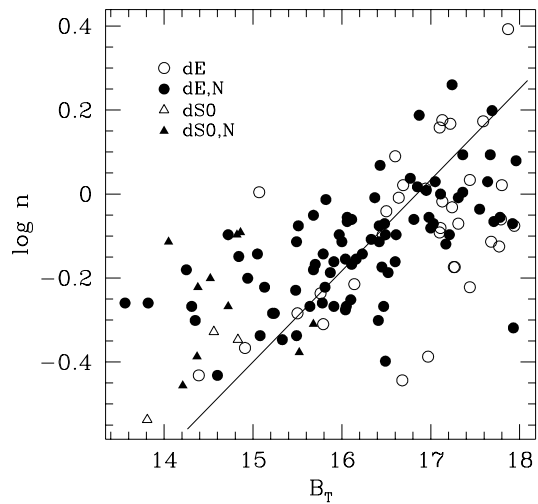}
\caption[]{The logarithm of S\'ersic's profile shape parameter $n$ versus total
apparent blue magnitude $B_{\rm T}$ 
for 128 early-type dwarf galaxies in the Virgo cluster.
Dwarf subtypes are plotted with different symbols, as given by the insert.
The straight
line, given by Eq.(6) in the text, is indicating the best-fitting linear 
correlation with $\log n$ as the independent variable.} 
\end{center}
\end{figure}
\begin{figure}
\begin{center}
\epsfxsize 75mm
\epsffile{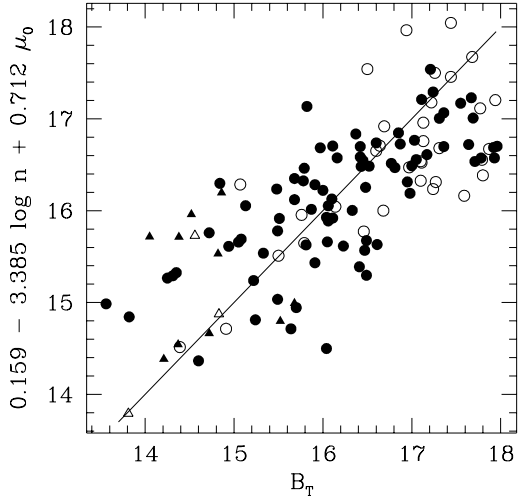}
\caption[]{The best-fitting linear combination of shape parameter $n$ and
central surface brightness $\mu_0$, given along the ordinate [= Eq.(8) in the
text], in a linear correlation with total magnitude $B_{\rm T}$. The line is 
identity. Sample and type coding are the same as in Fig.~3.}
\end{center}
\end{figure}

However, the correlation is significantly strengthened by plotting a linear
combination of $\log n$ and central surface brightness $\mu_0$ versus 
$B_{\rm T}$. The best-fitting combination, shown in Fig.~4, is
\begin{equation}
B_{\rm T} = - 3.385 \log n + 0.712\,\mu_0 + 0.159\,\,.
\end{equation}
Here the fitting was restricted to objects with $\log n \le$ 0 and/or
$\mu_0 \le$ 24 B\,arcesc$^{-2}$, again to avoid any bias due to
the magnitude cut-off at $B_{\rm T}$ = 18. The surface
brightness restriction will become plausible further
below (Fig.~5).  
The scatter has now reduced to $\sigma_{\rm obs} = 0.73$ mag. This means that
the {\em residuals}\/ of the $\log n - M$ relation are correlated with $\mu_0$.
As both $n$ and $\mu_0$ are distance-independent quantities, their combination
as given in Eq.(8) could well serve as distance indicator. But there is in 
fact no need to use a combination of these parameters. Surprisingly, 
the correlation of $\mu_0$ {\em alone}\/ with $B_{\rm T}$ is nearly as strong;
see Fig.~5. The linear regression here, restricted to $\mu_0 \le$ 23.5
B\,arcsec$^{-2}$, is
\begin{equation}
B_{\rm T} = 0.496\,\mu_0 + 5.426\,\,,
\end{equation}
with a scatter of $\sigma_{\rm obs}$ = 0.76 mag.
The fact that both $n$ and $\mu_0$ correlate with total magnitude must of 
course mean that $n$ and $\mu_0$ correlate with each other as well, which is
shown in Fig.~6. 

A scatter of 0.7 mag is also what one can get from the relation
between the mean {\em effective}\/ surface brightness 
$\langle \mu \rangle_{\rm eff}$ and total magnitude (BC91,
Jerjen 1995). The effective surface brightness has the great advantage to
be {\em model-independent}. So there is apparently no gain by the profile
fitting {\em with respect to the distance indicator application}. 

\begin{figure}
\begin{center}
\epsfxsize 75mm
\epsffile{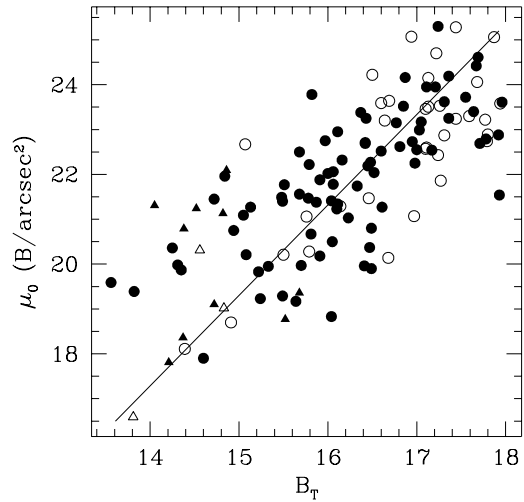}
\caption[]{S\'ersic's central surface brightness $\mu_0$ versus total 
magnitude $B_{\rm T}$. The linear regression line, with $\mu_0$ as independent
variable, is given by Eq.(9) in the text. Sample and type coding as in Fig.~3.}
\end{center}
\end{figure}
\begin{figure}
\begin{center}
\epsfxsize 75mm
\epsffile{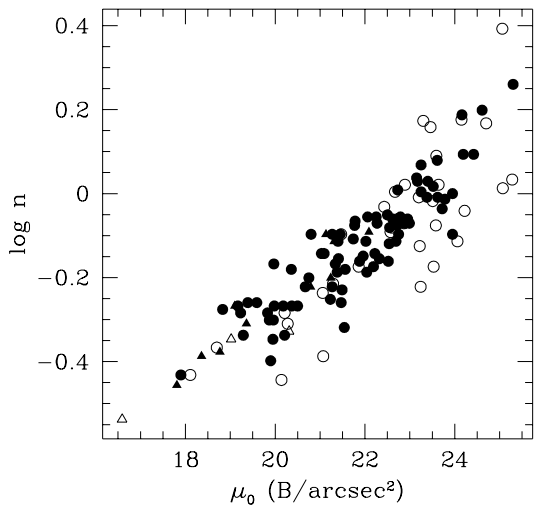}
\caption[]{The logarithm of profile shape parameter $n$ versus central surface
brightness $\mu_0$. Sample and type coding the same as in Fig.~3.}
\end{center}
\end{figure}

In their second paper, Young \& Currie (YC95) propose the $n - \log r$
relation as an alternative to $n$ versus $M$ as distance indicator. 
There is indeed a strong correlation between the two S\'ersic law 
parameters $n$ and $\log r_0$ as well, 
which is shown for our sample in Fig.~7. 
Again we use $\log n$ instead of $n$ as the independent quantity. The 
relation is certainly not linear. The best-fitting quadratic form is
\begin{equation}
\log r_0 = - 4.502\,(\log n)^2 + 1.876 \log n + 0.909\,\,,
\end{equation}
with an rms scatter in $\log r_0$ of $\sigma_{\rm log r} = 0.17$, which in
terms of magnitudes (by multiplying with 5) corresponds to $\sigma_{\rm m}$
= 0.85 mag. This is comparable to the scatter of the $n - M$ relation 
(Fig.~3).    

In contrast to the correlation with total magnitude (cf.~above), 
there is no improvement by replacing $\log n$ with $\mu_0$. This relation is
shown in Fig.~8. The linear regression line is
\begin{equation}
\log r_0 = 0.264\,\mu_0 - 5.292\,\,,
\end{equation}
with a large $\sigma_{\rm log r}$ = 0.25, or $\sigma_{\rm m}$ = 1.25 mag.
The scatter does not reduce by fitting a higher-order polynomial to the 
data.  
\begin{figure}
\begin{center}
\epsfxsize 75mm
\epsffile{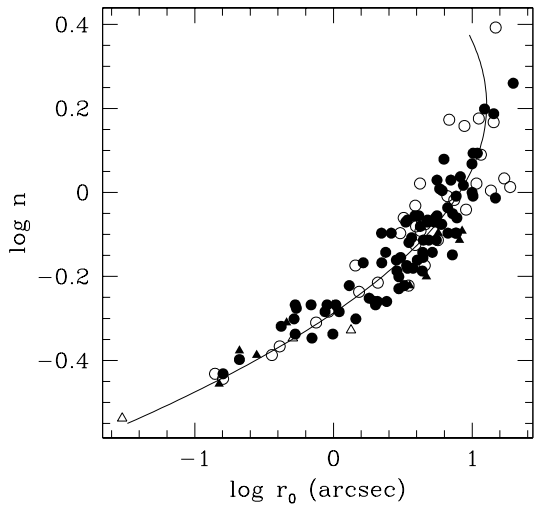}
\caption[]{The logarithm of shape parameter $n$ versus the logarithm of
S\'ersic's radius scale $r_0$. Sample and type coding are the same as in 
Fig.~3. The line is the best-fitting 2$^{\rm nd}$-order polynomial with 
$\log n$ 
as independent quantity, given by Eq.(10) in the text.}
\end{center}
\end{figure}
\begin{figure}
\begin{center}
\epsfxsize 75mm
\epsffile{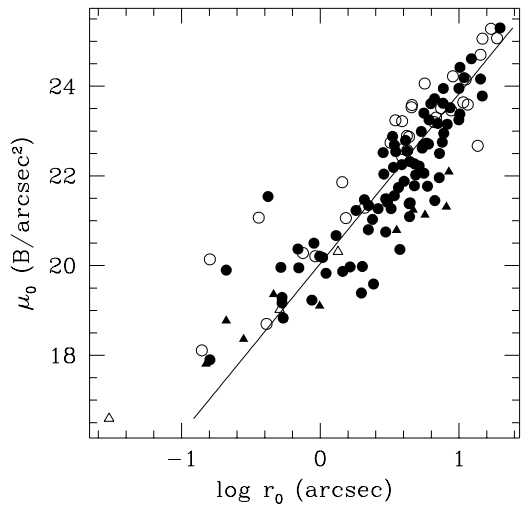}
\caption[]{S\'ersic's central surface brightness $\mu_0$ versus the logarithm
of radius scale $r_0$. Sample and type coding are the same as in Fig.~3.
The regression line is given by Eq.(11).}
\end{center}
\end{figure}

One may wonder what a scatter around the mean $\log n - \log r_0$ relation 
of only $\sigma_{\rm logr}$ = 0.17 really means, 
in view of the much larger
scatter in $\log r_0$ of $\sigma (\Delta \log r_0)$ = 0.40 for the difference
between the 
growth curve fits and the differential fits, and likewise 
$\sigma (\Delta \log r_0)$ = 0.32 for the difference between 
our $\log r_0$ values and
Durrell's (1997), as quoted in Sect.~2. Here, the surprising fact is
that the scatter in $\log r_0$ is again only $\sigma_{\rm logr}$ = 0.17, albeit
around a slightly different $\log n - \log r_0$ relation, if we switch
to differential profile fitting. However, if $n$ is determined from a 
growth curve fit and $\log r_0$ from a differential fit, or vice versa,
this scatter turns out to be $\sigma_{\rm logr} \approx$ 0.3. This we take
as evidence that the best-fitting S\'ersic parameters, for a given galaxy,
are {\em not independent}\/ from 
each other. We will elaborate on this important point further below.

Overall, then, the scatter in the S\'ersic profile scaling relations -- 
at least for our Virgo dwarf sample -- is unacceptably large for 
use in the distance determination business. At least, these relations 
would make only
rough \& unreliable distance indicators. We have also looked for any
dependence of the residuals on (1) the ellipticity of
the dwarf galaxies, and (2) the distance to the center of the cluster,
i.e.~M87. No trend was found. There is no way how the scatter could be reduced
to below $\simeq$ 0.7 mag. Thus the large scatter must essentially be 
{\em intrinsic -- unless the Virgo cluster has a much greater depth than
we thought}. This, in fact, is the claim of YC95.
It will be rejected in the following section.
\section{Critique of Young and Currie (1995)}
Young \& Currie's starting point (in YC94) 
was the $n - M$ relation for a sample
of 26 Fornax cluster E and dE galaxies, for which they had derived S\'ersic
law parameters from Caldwell \& Bothun's (1987) high-quality profiles. 
The scatter of this relation turned out to be a large 
$\sigma_{\rm m}$ = 0.88 mag.
Only by the exclusion of four ``outliers'' was the scatter reduced to an
encouragingly small $\sigma_{\rm m}$ = 0.47, and likewise to 
$\sigma_{\rm log r}$ = 0.108 for the $n - \log r$ relation (YC95).
One could question this procedure, and one could criticize the obvious
incompleteness of this Fornax cluster sample. But this is of no concern here.
If the large scatter for Virgo cluster dwarfs is truly intrinsic, the 
usefulness, or reliability of these scaling relations as distance indicators
is undermined at once -- no matter what the reason for this difference
between Fornax and Virgo is.

Our concern is Young \& Currie's claim that the intrinsic scatter for 
Virgo dwarfs is as small as for Fornax dwarfs, which would mean that
the large observed
scatter for Virgo must be a {\em depth effect}. Such a claim can only
be based on some {\em independent}\/ information on the intrinsic scatter.
YC95 apparently seized this information by applying both the $n - M$ and the 
$n - \log r$ relation {\em at the same time}.

Let us first see what happens if we do this with our own data. In Fig.~9
we have plotted the residuals from 
the $\log n - B_{\rm T}$ relation, calculated
with respect to the regression line given by Eq.(6), {\em versus}\/ $5\times$
the residuals from the $\log n - \log r_0$ relation, 
calculated with respect
to the quadratic form given by Eq.(10). Obviously, the residuals are 
correlated. A formal regression for a fixed slope of 1, i.e.
\begin{equation}
{\rm res} (B_{\rm T} - \log n) = - 5 \cdot {\rm res} 
(\log r_0 - \log n)\,\,,
\end{equation}
gives a scatter of $\sigma_{\rm m}$ = 0.86 mag, 
which is comparable to the values of
$\sigma_{\rm m}$ = 0.92 (Fig.~3) and 0.85 mag 
(Fig.~7), respectively, for either relation {\em alone}. This must mean
that the two relations, i.e.~the parameters involved, are 
{\em not independent} -- a suspicion spelled out already in the 
preceding
section. The argument is quite simple: -- {\em if}\/ we would deal with two 
independent but equivalent
``distance measurements'' (in YC's terms),
the scatter of the difference between the residuals, i.e.~``distances''
should equal the scatter of their sum, i.e.~it should become {\em larger}\/ 
than the scatter of a single relation by $\approx \sqrt[]{2}$ (assuming
Poissonian statistics).

Such an interdependence
should indeed be expected, as the three profile
parameters of S\'ersic's generalysed law are connected with total magnitude
via Eq.(5): only three of the four quantities involved can be free.
Our claim is that {\em the mean relation between $\log n$ and $\mu_0$}\/
(Fig.~6) {\em is sufficient 
to cause the residual correlation}\/ (Fig.~9). To prove this we have 
performed a simple Monte Carlo calculation.

The procedure was as follows: for the same dwarf sample we took the observed
$B_{\rm T}$ and the best-fitting $\log n$ values. For every galaxy we then
determined $\mu_0$ from a Gaussian random distribution around the 
mean $\log n - \mu_0$ relation (Fig.~6) with $\sigma_{\mu_0}$ = 0.83
mag (note that both of these parameters are distance independent,
hence the cluster depth cannot sneak in here). The forth parameter, $\log r_0$,
is then fixed by Eq.(5), assuming perfect profile fits. Once all parameters
are given, we fitted a quadratic form to the $\log n - \log r_0$ relation
and finally determined the scatter of the relation
between the $B_{\rm T} - \log n$ and $\log r_0 - \log n$ residuals, i.e.~the
scatter of the residual differences.
This was repeated 10\,000 times, after which everything was averaged. 
The average $\log n - \log r_0$ relation turned out to be very similar 
to the observed one, though with a somewhat larger scatter of
$\sigma_{\rm logr}$ = 0.24 (versus 0.17 as observed), which corresponds to
$\sigma_{\rm m}$ = 
1.2 mag. Notwithstanding, the average scatter of the difference between
res($B_{\rm T} - \log n$) and $-5 \cdot {\rm res}(\log r_0 - \log n$) is
again as low as $\sigma$ = 0.85 mag (versus 0.86 as observed). For comparison,
the scatter of their sum is 1.94 mag. 

It is easy now to see what happens if this {\em intrinsic}\/ connection
between the $\log n - B_{\rm T}$ and $\log n - \log r_0$ relations is 
neglected: If we {\em assume}\/
with Young \& Currie (1995)
that ``... the $L - n$ and $R - n$ estimates are,
to a good approximation, independent'', we will {\em divide}\/ 
$\sigma_{\rm m}$ = 0.86 mag from above 
by (at least) $\sqrt[]{2}$ and arrive at an 
``intrinsic'' scatter of 0.61 mag for a single relation. 
Since this is significantly
smaller than the
0.92 mag scatter of the $\log n - M$ relation, we are forced to 
Young \& Currie's cluster depth interpretation, with 
$\sigma_{\rm depth} \approx$ 0.6 mag in this case, if we allow also 
for observational 
errors (cf.~Sect.~3).

\begin{figure}
\begin{center}
\epsfxsize 75mm
\epsffile{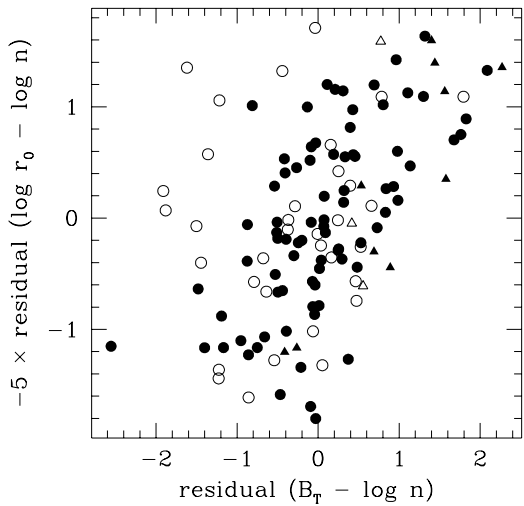}
\caption[]{Residual from the $\log n - B_{\rm T}$ relation [Eq.(6)] versus
five times the residual from the $\log n - \log r_0$ relation [Eq.(10)]. 
The quantities plotted are magnitudes. Sample and type coding are the same as
in Fig.~3.}
\end{center}
\end{figure}

However, this does not yet explain why YC95 found a very small 
``intrinsic'' scatter
of 0.4 - 0.5 mag 
for their Virgo dwarfs. We think the reason for this additional effect 
must lie in the use of low-resolution (Schmidt) plates for surface photometry, 
where, e.g., the semi-stellar, central
nuclei are smeared into the general profiles of the galaxies. 
In order to achieve a
comparison with their high-resolution calibrating sample of Fornax cluster
and Local Group dwarfs, YC95 had to convolve the profiles of the
latter with a broad (FWHM $\approx 5\arcsec$) Gaussian, upon which the
best-fitting S\'ersic law parameters were determined. This was done for
a whole grid of distance moduli of the calibrating galaxies. Then, for each
individual Virgo dwarf a distance modulus was read from its locus with
respect to the calibrating grid in the $n - M$ and $n - \log r$ plane,
respectively. Finally, the comparison of the individual moduli as determined
in both planes yielded the small ``intrinsic'' dispersion mentioned above.
Here our suspicion is that {\em this small scatter merely mirrors the small 
scatter of the calibrating
dwarfs}, having nothing to do with the Virgo sample. The convolution process
changes the calibrating profiles {\em coherently}\/ for both the $n - M$
and $n - \log r$ relations, such that the difference in distance for the
Virgo dwarfs from these relations is bound to collapse to the intrinsic
dispersion of the calibrating sample. 

Whether our suspicion is correct could only be tested by a complete
simulation of the analysis of YC95. However, as we have shown -- based on
the present data of superior quality -- that the
assumption of independence between the $n -M$ and $n - \log r$ relations
is fundamentally flawed, there is no need for such a pursuit.
In the absence of any independent evidence to the contrary, 
it is a principle of {\em conceptual economy}\/ to assume
that the large scatter of the Virgo sample is {\em intrinsic}, especially as 
long as there is no comparison sample (e.g., Fornax cluster) of similar
completeness.    

Having argued that there is no evidence that the large intrinsic scatter of 
Virgo sample is {\em not}\/ intrinsic, we can also ask -- Is there any
positive evidence that it {\em is}\/ intrinsic? We think the following simple 
test does provide such evidence. 

Suppose the observed large scatter of the Virgo dwarfs {\em is}\/ a depth 
effect, and the dwarfs are distributed in a prolate structure (rather than in
a compact core as we thought) -- pointing towards us (never mind!) and
stretching from 8 to 20 Mpc in YC's distance scale (see Fig.~4 in YC95).
Then these dwarfs (outside a cluster core) would have to possess rather 
small peculiar velocities, of order 50 km s$^{-1}$ or less, and we would
{\em expect to find a well-defined velocity-distance relation}\/ for them,
which should be the Hubble flow modulo a virgocentric infall pattern.

Fortunately, a fairly large subsample of 43 objects with known velocities
out of our 128 Virgo dwarfs makes this test feasible. The velocities 
are listed in Table 1 (data from Binggeli et al.~1985, 1993).
On the above assumption we now calculate pseudo distances for these 43 dwarfs
from the residuals of the $\log n - B_{\rm T}$ relation:
\begin{equation}
D_{\rm pseudo} = 20 \cdot 10^{- 0.2\,{\rm res}\,(B_{\rm T} - \log n)}\,\,
[{\rm Mpc}]\,\,,
\end{equation}
where an average distance of 20 Mpc was chosen. It should be mentioned that
in this case the regression line to which
the residuals refer was calculated only for those 43 dwarfs; 
it is slightly,
but not significantly different from Eq.(6).

In Fig.~10 we have plotted
the resulting pseudo distances versus the heliocentric 
velocities for the 43 Virgo
dEs with known redshifts. The same is shown in Fig.~11 for the better defined
$\mu_0 - B_{\rm T}$ relation. 

Obviously, our Virgo dwarfs do not follow a well-behaved velocity-distance
relation. In the absence of any intrinsic scatter, all points should lie
{\em within}\/ the ``infall boundary'' indicated by the curved lines. These
lines give the loci of galaxies that are falling into the cluster center
along our line of sight (hence marking the {\em maximum}\/ radial velocity).
They are calculated with Kraan-Korteweg's (1986) infall model for a LG infall
velocity of 220 km s$^{-1}$. Even with an intrinsic distance uncertainty of,
say, 20 \% (following YC95), i.e.~$\sigma_{\rm D} \approx$ 4 Mpc, or 
$\sigma_{\rm m} \approx$ 0.4 mag, there are simply 
too many dwarfs (30 \%)
lying outside these boundaries to comply with Young \& Currie's hypothesis.

Especially troublesome are the dwarfs with {\em negative}\/ velocities. There
are a number of them (5 out of $<$ 100 with known redshifts), some  
of which have $v = - 700$ km s$^{-1}$ ! 
(cf.~Binggeli et al.~1993). It is clear -- for dynamical reasons -- that
these objects {\em must}\/ lie in the core of the Virgo cluste.

It is, in fact, the
mere {\em existence}\/ of a small number of negative velocities among the 
dwarfs which provides the strongest (and most simple!) argument against the
cluster depth hypothesis. For if these dwarfs are in the core (as admitted), 
their velocities most likely consitute the low-velocity tail of a broad 
(more or less Gaussian) velocity distribution.
But this implies that for every dwarf
with a negative velocity, there must be {\em many more}\/ dwarfs -- 
occupying the same
spatial area, i.e.~the cluster core -- with a positive radial velocity. 
Hence, by a qualitative statistical argument, the {\em majority}\/ 
of early-type 
Virgo dwarfs 
must certainly lie in the cluster core.

\begin{figure}
\begin{center}
\epsfxsize 80mm
\epsffile{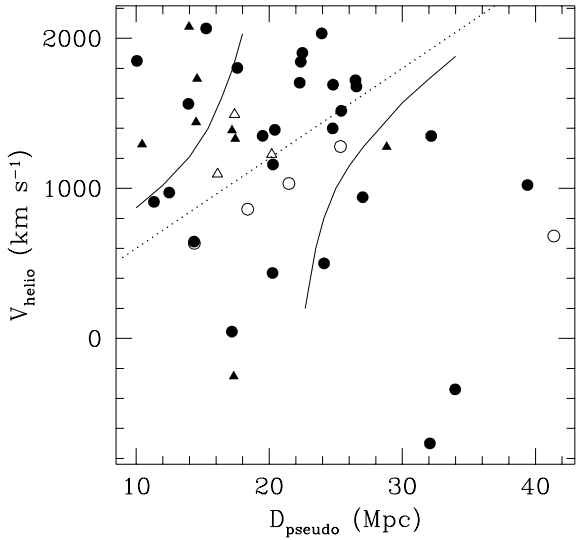}
\caption[]{Heliocentric velocity versus pseudo distance calculated from
the $\log n - B_{\rm T}$ residual according to Eq.(13) for 43 Virgo 
dEs and dS0s with known redshifts. 
The type coding is the same as in Fig.~3.
The dotted line corresponds to a quiet Hubble flow with $H_0$ = 60 
km s$^{-1}$ Mpc$^{-1}$. A galaxy falling through the cluster center ($D$ = 20
Mpc, $\langle v \rangle$ = 1200 km s$^{-1}$) along our line of sight would
move on one of the two curved lines. These are based on the virgocentric infall
model of Kraan-Korteweg (1986) with a LG infall velocity of $v_{\rm LG}$ =
220 km s$^{-1}$.}
\end{center}
\end{figure}
\begin{figure}
\begin{center}
\epsfxsize 80mm
\epsffile{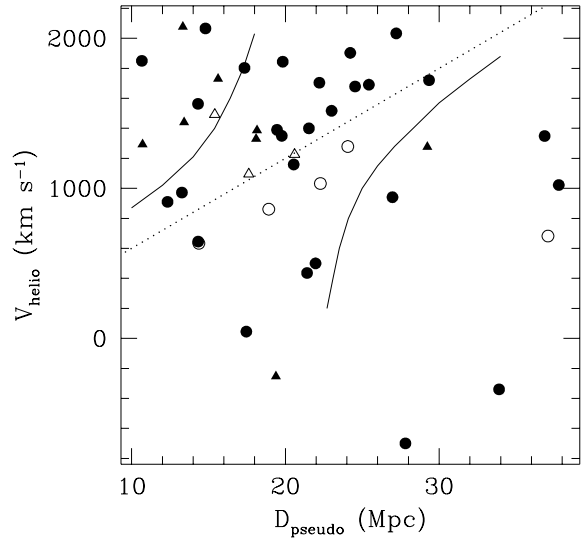}
\caption[]{The same as Fig.~10 but for the residuals of the 
$\mu_0 - B_{\rm T}$ relation [Fig.~5, Eq.(9)].}
\end{center}
\end{figure}
\section{Conclusions}
There is no evidence that the large scatter of the S\'ersic profile 
parameter scaling relations for Virgo cluster dEs, 
in particular shape parameter $n$ versus
$M$ or $\log r$, is other than intrinsic, i.e.~being due to a (unexpectedly)
large depth of the Virgo cluster. Young \& Currie's (1995) claim to the
contrary is based on low-resolution photometry and the untenable assumption
that the $n - M$ and $n - \log r$ relations are independent. 
On the other hand, there is some evidence in favour of a large
intrinsic scatter by the absence of 
any correlation between the radial velocity and the residual from the 
$n - M$ relation for a sample of 43 Virgo dwarfs with known redshifts. 
Such a correlation would be expected if the $n - M$ residual were 
essentially a measure for the distance of a galaxy.

In consequence, 
we need not revise our view of the spatial distribution of
dE galaxies and the morphology-density relation for dwarfs in general
(e.g.~Ferguson \& Binggeli 1994). Dwarf ellipticals loosely distributed in a
prolate cloud structure as put forward by YC95 would have been very difficult
to understand, given that dEs are not known to exist (except as close 
companions) in the nearby ``cloudy'' field. The broad velocity distribution
of Virgo cluster dEs, and the well-populated tail of negative velocity 
dwarfs in particular, is independent evidence that these galaxies are lying
in the deep potential, i.e. in the narrow space of the core of the cluster.
The filamentary distribution of Virgo spirals and irregulars (Fukugita et 
al.~1993, Yasuda et al.~1997), which was mentioned by YC95 as corroborative
evidence, is an entirely different story. These types of galaxies are
known to {\em avoid}\/ the cluster core (Binggeli et al.~1987).

If the intrinsic dispersion of the $n - M$ or $n - \log r$ relation is much
smaller for Fornax dwarfs than for Virgo dwarfs as it appears (which,
however, might be caused by the incompleteness of YC's Fornax sample),
we are in need of an explanation for this difference. In any case, the large 
intrinsic scatter for Virgo dwarfs is certainly not in favour of an
application of the S\'ersic profile scaling relations to distance 
measurements. These relations are very interesting and important in the
context of the connection between E and dE (Jerjen \& Binggeli 1997),
but they can only be of limited use as distance indicators: the uncertainty
in the distance modulus for a single galaxy cannot be reduced to below
$\approx$ 0.7 mag. This is clearly inferior to the Tully-Fisher and
$D_{\rm n} - \sigma$ methods (e.g. Jacoby et al.~1992), and it is no better
than what can be achieved with the $\langle \mu \rangle_{\rm eff} - M$
relation for the same galaxies, which does not involve any profile fitting.

\begin{acknowledgements}
Financial support by the Swiss National Science Foundation is gratefully
acknowledged. We thank Dr.~J.~van Gorkom for a crucial hint, and the referee,
Dr.~M.Fukugita, for his critical comments which helped to improve this paper.
\end{acknowledgements}

\end{document}